\documentclass[
	aps, aps,prd,longbibliography, superscriptaddress, twocolumn,
	10pt
	floatfix, 
    nofootinbib,
	tightenlines
]{revtex4-1}
\usepackage[final]{graphicx}
\usepackage{times,bbm,amsmath,amssymb}
\usepackage{epsfig,color}
\usepackage{xcolor}
\usepackage{hyperref}
\hypersetup{
    colorlinks = true
}
\usepackage{cleveref}
\usepackage{microtype}

\usepackage{float,siunitx}
\usepackage[caption = false]{subfig}

\usepackage[greek,english]{babel}
\usepackage{thumbpdf,enumerate}
\usepackage{booktabs}
\usepackage{sidecap}
\usepackage[scaled=.8]{couriers}
\usepackage{multirow}
\usepackage{placeins}
\usepackage{relsize}
\usepackage{pst-grad,bm}
\usepackage{epigraph}
\usepackage{gensymb}
\usepackage{longtable}
\usepackage{ulem} 
\usepackage{acronym}
\usepackage{physics}
\usepackage{easyReview}

\usepackage[columnwise]{lineno}

\DeclareUnicodeCharacter{0301}{\'{e}}

\begin{document}


\vspace{10pt}


\title{Biphoton State Reconstruction via Phase Retrieval Methods}

\author{Nazanin Dehghan}
\address{Nexus for Quantum Technologies, University of Ottawa, Ottawa, K1N 6N5, ON, Canada}
\affiliation{National Research Council of Canada, 100 Sussex Drive, K1A 0R6, Ottawa, ON, Canada}

\author{Alessio D'Errico} 
\email{aderrico@uottawa.ca}
\address{Nexus for Quantum Technologies, University of Ottawa, Ottawa, K1N 6N5, ON, Canada}
\affiliation{National Research Council of Canada, 100 Sussex Drive, K1A 0R6, Ottawa, ON, Canada}

\author{Francesco Di Colandrea} 
\address{Nexus for Quantum Technologies, University of Ottawa, Ottawa, K1N 6N5, ON, Canada}

\author{Ebrahim Karimi}
\address{Nexus for Quantum Technologies, University of Ottawa, Ottawa, K1N 6N5, ON, Canada}
\affiliation{National Research Council of Canada, 100 Sussex Drive, K1A 0R6, Ottawa, ON, Canada}

\begin{abstract}
The complete measurement of the quantum state of two correlated photons requires reconstructing the amplitude and phase of the biphoton wavefunction. We show how, by means of spatially resolved single photon detection, one can infer the spatial structure of bi-photons generated by spontaneous parametric down conversion. In particular, a spatially resolved analysis of the second-order correlations allows us to isolate the moduli of the pump and phasematching contributions to the two-photon states. When carrying this analysis on different propagation planes, the free space propagation of pump and phasematching is observed. This result allows, in principle, to gain enough information to reconstruct also the phase of pump and phasematching, and thus the full biphoton wavefunction. We show this in different examples where the pump is shaped as a superposition of orbital angular momentum modes or as a smooth amplitude with a phase structure with no singularities. The corresponding phase structure is retrieved employing maximum likelihood or genetic algorithms. These findings have potential applications in fast, efficient quantum state characterisation that does not require any control over the source.
\end{abstract}

\maketitle 
\section{Introduction}
\begin{figure*}
\centering
\includegraphics[width=1\textwidth]{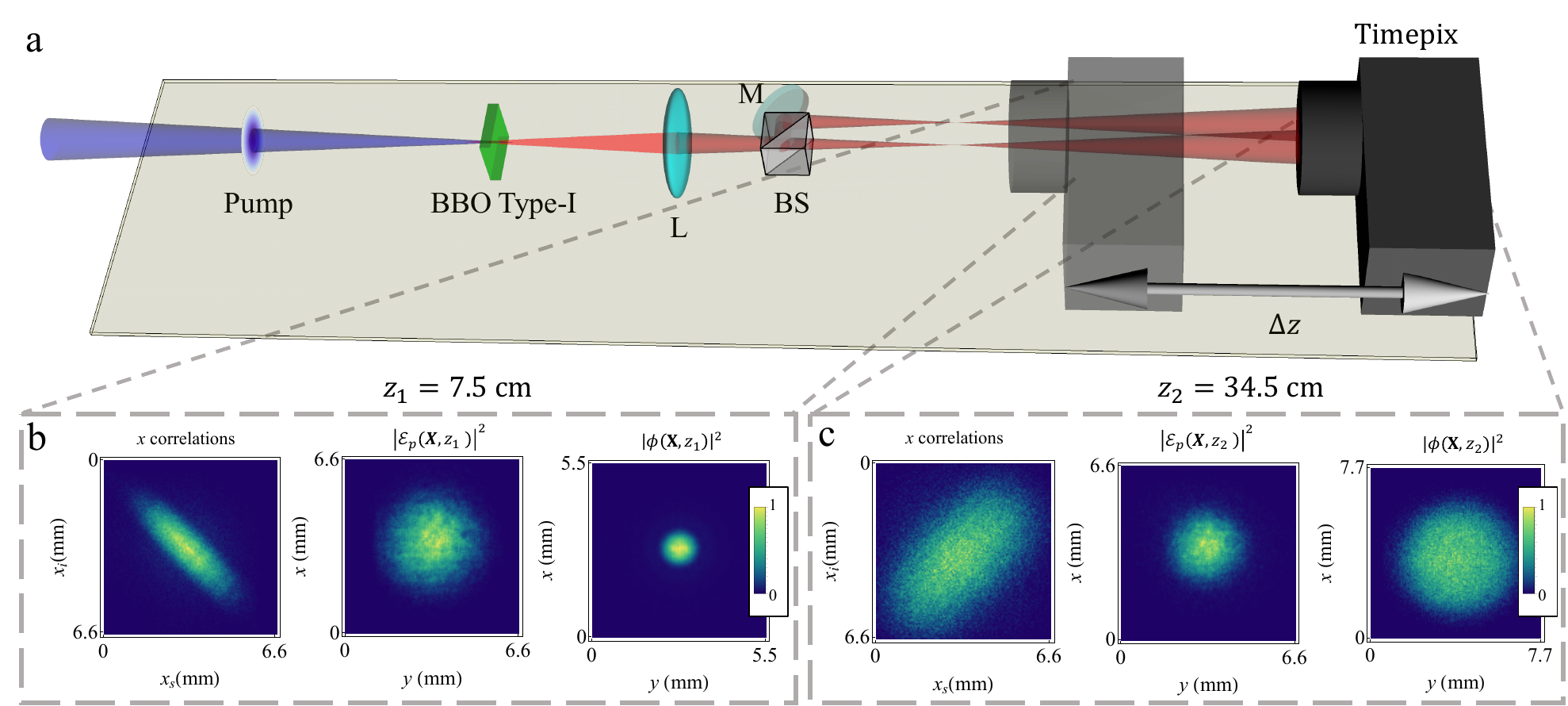}
\caption{\textbf{Experimental layout and coincidence analysis.} a- Simplified sketch of the experimental setup. A pump laser (central wavelength 405 nm) is incident on a 0.5 mm thick BBO Type-I crystal that generates photon pairs with the same polarisation. The photons propagate through an imaging system (depicted here as a single lens L; see Methods for the detailed experimental layout). Signal and idler photons are spatially separated with a 50/50 beamsplitter (BS) and impinge on a time-stamping camera (Tpx3D), which allows for retrieving spatially resolved coincidence counts. The camera was moved along the propagation direction to measure spatial correlations in different planes and extract the corresponding pump and phasematching function intensities. An example with a Gaussian pump is shown in panels b and c.}
\label{fig:setup}
\end{figure*}

\begin{figure} 
\includegraphics[width=1\columnwidth]{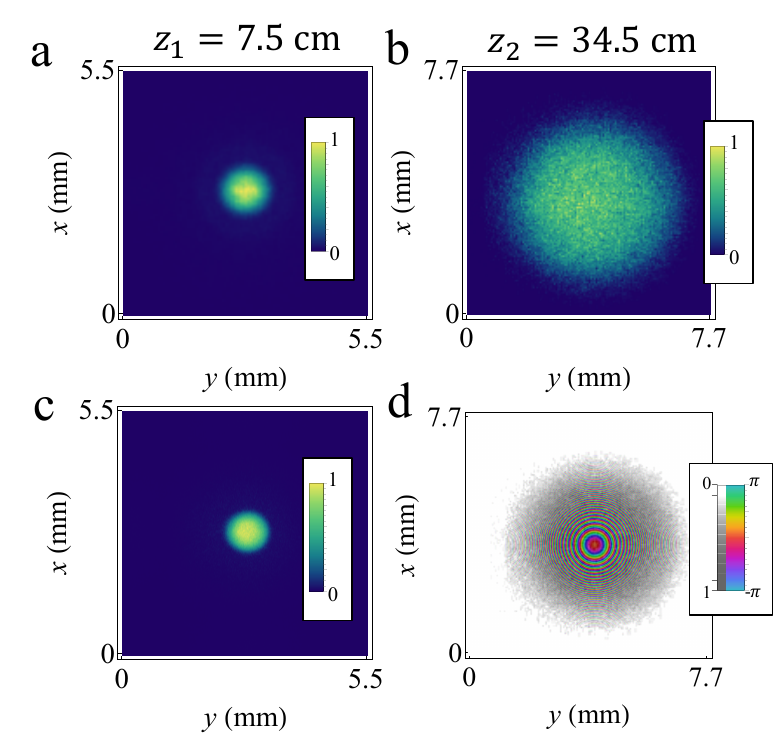}
\caption{\textbf{Full reconstruction of phasematching function.} a-b. Experimentally extracted intensity of the phasematching function of an SPDC state in two different propagation planes. c. Reconstructed phasematching intensity in $z_1$ assuming the quadratic phase shown in panel d and applied on the experimental amplitude in $z_2$. The intensity in $z_1$ is obtained by Fresnel propagation. The excellent agreement with the experimental intensity in $z_1$ indicates the correctness of the assumed phase structure. }
\label{fig:phasematching}
\end{figure}
The tomography of quantum states plays a fundamental role in modern quantum technologies~\cite{ james2001measurement,d2003quantum,torlai2018neural, rambach2021robust, struchalin2018adaptive}. At the same time, such a task can be particularly challenging when dealing with systems of many particles and/or many degrees of freedom \cite{agnew2011tomography,d2021full}. If a projective measurement approach is adopted, with no prior information, the number of required measurements scales quadratically with the dimensionality of the Hilbert space \cite{james2001measurement}. 
The case of the reconstruction of high dimensional two-photon states is of particular interest given their applications in fundamental quantum mechanics~\cite{dada2011experimental,malik2016multi,chen2017experimental, cervera2022experimental}, high-dimensional quantum communications~\cite{watanabe2008tomography,bouchard2019quantum}, and quantum imaging~\cite{genovese2016real}.
As an alternative to projective approaches based on mode sorting, recent works introduced the possibility of exploiting variations of classical interferometric techniques for directly reconstructing the spatial structure of phase and amplitude of an unknown biphoton state \cite{zia2023interferometric}. This approach is practically faster and more reliable than traditional methods thanks to the use of time-stamping cameras~\cite{nomerotski2019imaging,morimoto2020megapixel, nomerotski2023intensified}, which is proving to be a promising resource for quantum optics experiments~\cite{defienne2021polarization, zhang2021high,morimoto2021superluminal,zhang2022ray,ndagano2022quantum, defienne2022pixel,zia2023interferometric,thekkadath2023intensity,wang20243d,Zhang2024}. However, interferometric approaches will require the phase locking of the reference biphoton state with the unknown one, a task which can be harder to achieve in cases where the unknown source is not easily accessible. It is thus desirable to have the possibility of obtaining the phase structure of the unknown biphoton from measurements that do not require any control of the source. Here, we propose a method based on measuring the spatial coincidence distribution in different propagation planes and reconstructing the phase of the biphoton employing phase retrieval algorithms. 

We demonstrate how the analysis of coincidence images allows us to extract the square moduli, hereafter referred to as ``intensities'', of the two complex functions that contribute to the biphoton state: the pump and the phasematching. Intriguingly, this separation can be performed in any propagation plane (assuming free space propagation and eventual lenses that can be present in the setup). The intensity of the extracted pump or phasematching function in a given plane is related to the respective fields in another plane at a distance $d$ by paraxial propagation through a distance $d/2$, or assuming the pump and phasematching functions to have half the wavelength of the biphoton state. Retrieving pump and phasematching intensities at different planes and knowing their relationship opens the possibility of reconstructing the phase of these fields and, thus, the full biphoton wavefunction. We demonstrate theoretically and experimentally how to isolate the pump and phasematching intensities from coincidence images, and we give some examples of their phase reconstruction based on the exploitation of maximum likelihood approaches --in the case a decomposition in orthogonal mode sets can be conveniently used-- or genetic algorithms -- in the case of aberrated pumps with smooth amplitudes--.

\section{Theory}
Consider a two-photon state with fixed wavelengths and polarisations, in a paraxial propagation regime. In this case, assuming the $z$-axis corresponding to the mean propagation direction, the system can be described in terms of the probabilities of photons passing through the transverse position $\mathbf{X}=(x,y)$ in a given plane. The quantum state is thus described in terms of the creation operators $\hat{a}^{\dagger}_{\mathbf{X}}$. Defining the transverse position eigenstate as $\ket{\mathbf{X}}:=\hat{a}^{\dagger}_{\mathbf{X}}\ket{0}$, with $\ket{0}$ the vacuum state, a general biphoton state is given by
\begin{align}
    \ket{\Psi}=\iint \Psi(\mathbf{X}_i,\mathbf{X}_s)\ket{\mathbf{X}_i}\otimes\ket{\mathbf{X_s}}\,d\mathbf{X}_i\,d\mathbf{X}_s,
\label{eq:biphoton}
\end{align}
where $\Psi(\mathbf{X}_i,\mathbf{X}_s)$ is called the biphoton wavefunction. The indexes $i$ and $s$ denote, conventionally, \textit{idler} and \textit{signal} photons, respcetively. Equation~\eqref{eq:biphoton} is a representation of the state in a basis of transverse coordinates eigenstates for a given value of $z$. The free space propagation of the biphoton wavefunction from $z$ to $z'$ is 
\begin{align}
    \Psi'(\mathbf{X'}_i,\mathbf{X'}_s)=\iint&\mathcal{G}_{\Delta z}(\mathbf{X'}_i,\mathbf{X}_i)\mathcal{G}_{\Delta z}(\mathbf{X'}_s,\mathbf{X}_s)\times\cr&\Psi(\mathbf{X}_i,\mathbf{X}_s)d\mathbf{X}_i\,d\mathbf{X}_s,
\label{eq:biphotonpropagator}
\end{align}
where $\mathcal{G}_{\Delta z}(\mathbf{X'},\mathbf{X})$ is a free space propagator through the distance $\Delta z=z'-z$ and $\mathbf{X'}=(x',y')$ are the transverse coordinates in the plane $z'$. For paraxial fields, the free space propagation is described in terms of the Fresnel propagator
\begin{align}
    \mathcal{G}_z(\mathbf{X}',\mathbf{X})=\frac{i}{\lambda z}e^{ikz}\exp(-i\frac{\pi}{\lambda z}(\mathbf{X}'-\mathbf{X})^2).
\end{align}

A common and interesting case is correlated biphoton states with the following wavefunction,
\begin{equation}
    \Psi(\mathbf{X}_i,\mathbf{X}_s)=\mathcal{E}_p\biggr(\frac{\mathbf{X}_i+\mathbf{X}_s}{2}\biggr)\,\phi\biggr(\frac{\mathbf{X}_i-\mathbf{X}_s}{2}\biggr).
\end{equation} 
In particular, in the case of biphotons generated via spontaneous parametric down-conversion (SPDC) in Type-I crystals, $\mathcal{E}_p$ is the spatial, slowly varying amplitude of the pump laser, and $\phi$ is the phasematching function~\cite{walborn2010spatial,miatto2012spatial}. This structure simplifies the propagation of the biphoton wavefunction significantly. Performing the change of variables $\mathbf{R}=(\mathbf{X}_i+\mathbf{X}_s)/2$, $\mathbf{\Delta}=(\mathbf{X}_i-\mathbf{X}_s)/2$ \cite{defienne2022pixel}, and expanding the squares $(\mathbf{X}'-\mathbf{X})^2$ in the Fresnel propagator, it is straightforward to verify that:
\begin{equation}
    \mathcal{G}_z(\mathbf{X}'_i,\mathbf{X}_i)\mathcal{G}_z(\mathbf{X}'_s,\mathbf{X}_s)=\frac{1}{4}e^{ikz}\mathcal{G}_{z/2}(\mathbf{R}',\mathbf{R})\mathcal{G}_{z/2}(\boldsymbol{\Delta}',\boldsymbol{\Delta}).
\end{equation}

This allows one to separate the 4-dimensional (4D) integral describing the SPDC propagation in free space into the product of two bi-dimensional integrals:
\begin{eqnarray}
    \Psi'(\mathbf{R}',\mathbf{\Delta}')=&\mathcal{N}&\int \mathcal{G}_{\Delta z/2}(\mathbf{R}',\mathbf{R})\mathcal{E}_p(\mathbf{R})d\mathbf{R}\cr
    &\times&\int \mathcal{G}_{\Delta z/2}(\mathbf{\Delta}',\mathbf{\Delta})\phi(\mathbf{\Delta})d\mathbf{\Delta},
\end{eqnarray}
where $\mathcal{N}$ is a normalisation constant. Thus, the propagated biphoton wavefunction has the structure $\Psi'(\mathbf{R}',\mathbf{\Delta}')=\mathcal{E}'_p(\mathbf{R}')\phi'(\mathbf{\Delta}')$, where $\mathcal{E}'_p$ and $\phi'$ are obtained applying a single Fresnel propagator on the functions $\mathcal{E}_p$ and $\phi$. It is important to emphasize that the propagation of $\mathcal{E}_p$ and $\phi$ must be evaluated for $\Delta z/2$ if $\Delta z$ is the propagation distance considered for the biphoton state. 

Experimentally, from a spatially resolved coincidence detection, one can extract the pump and phasematching contributions in each measurement plane. Coincidence detection allows to retrieve the 4D function $\mathcal{C}(\mathbf{X}_i,\mathbf{X_s})=\abs{\Psi(\mathbf{X}_i,\mathbf{X_s})}^2$. From this data set, the pump and phasematching squared moduli can be extracted when post-selecting on spatially correlated and anticorrelated states, respectively. This is done by evaluating the quantities
\begin{align}
    \mathcal{C}(\mathbf{X}_i+2\mathbf{c},\mathbf{X}_i)=\abs{\mathcal{E}_p(\mathbf{X}_i+\mathbf{c})}^2\abs{\phi(-\mathbf{c})}^2,
\label{pump}
\end{align}
and
\begin{align}
    \mathcal{C}(-\mathbf{X}_i+2\mathbf{c},\mathbf{X}_i)=\abs{\mathcal{E}_p(\mathbf{c})}^2\abs{\phi(\mathbf{X}_i-\mathbf{c})}^2.
\label{phasematch}
\end{align}
Apart from a constant shift $\mathbf{c}$, the post-selected coincidences give patterns that are spatially distributed as the pump and the phasematching intensity.
Here, we considered a generic case in which correlations and anti-correlations are chosen with a constant offset $\mathbf{c}$. This allows for increasing the statistics without the need for long exposure times in the experiment -- see the following section and Supplementary Figure S1.

Using the technique described, we can extract the individual functions of the pump $\mathcal{E}_p$ and phasematching $\phi$ that contribute to the biphoton state. We can then observe how these functions propagate through free space. By examining the distributions of the pump and phasematching functions in different planes, it is possible to obtain information on their phase structure without the need for interferometric or tomographic measurements. \newline

\section{Experimental Results}
\begin{figure*}
\centering
\includegraphics[width=1\textwidth]{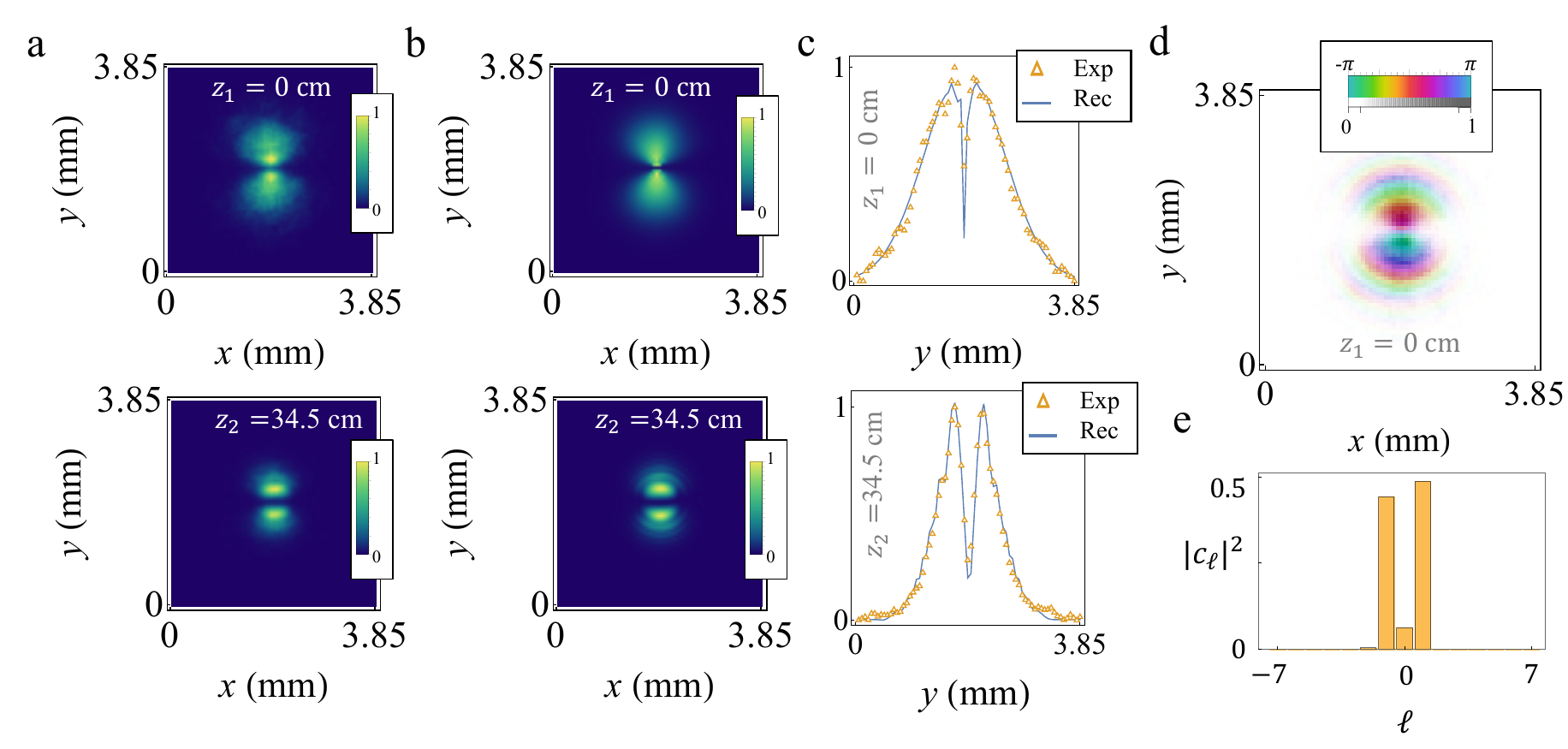}
\caption{\textbf{Phase reconstruction of the anti-symmetric state.} a.~Experimentally extracted pump intensities for an anti-symmetric SPDC state. b,c. Reconstructed intensities from an optimal superposition of HyGG modes. Panel c shows the comparison with the experiment (gold-coloured triangles) along the line $x=1.6$ mm. d.~Reconstructed phase and amplitude of the pump contribution in $z_1$. e.~The OAM power spectrum of the pump, which was obtained from the reconstruction.}
    \label{fig:hg10}
\end{figure*}
\begin{figure}
\centering
\includegraphics[width=1\columnwidth]{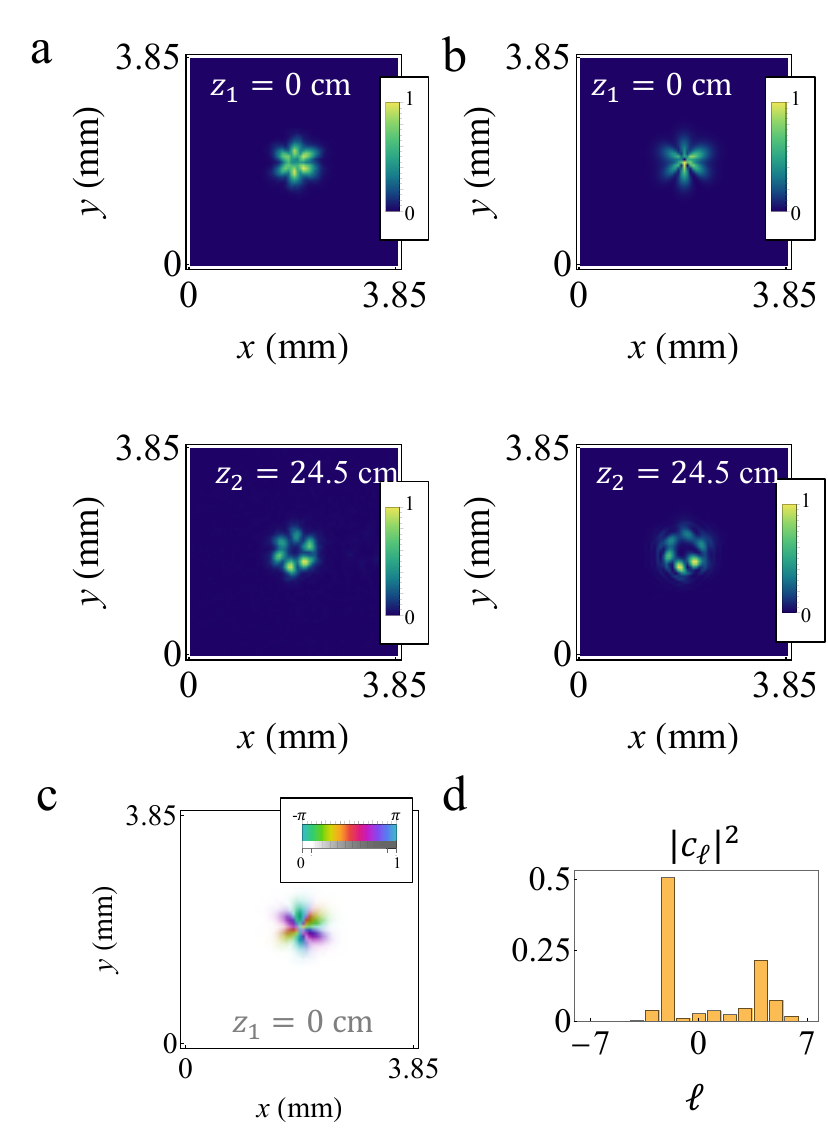}
\caption{\textbf{Phase reconstruction of OAM superposition states.} a.~Experimentally retrieved pump intensity in a superposition of, nominally, OAM=4 and OAM=-2. b.~Intensities of the reconstructed field obtained by superimposing 15 HyGG modes. c.~Retrieved phase and amplitude of the pump contribution at the crystal image plane. d.~The OAM power spectrum of the reconstructed field.}
    \label{fig:oam4_2}
\end{figure}
\begin{figure*}
\centering
\includegraphics[width=1\textwidth]{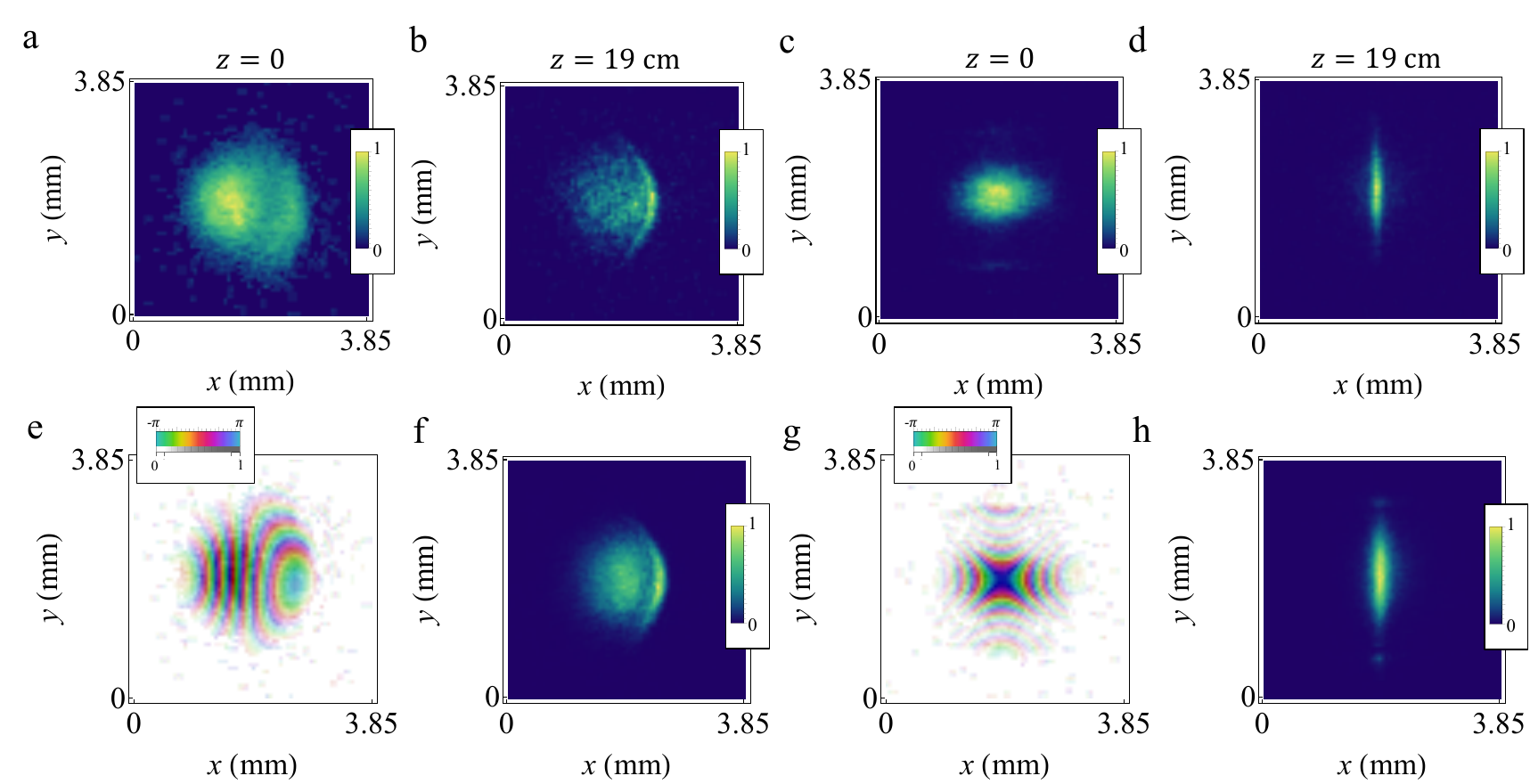}
\caption{\textbf{Phase reconstruction of aberrated Gaussian pumps.}  Panels a to d show the extracted pump shapes in two propagation planes, $z=0$ and $19$ cm. For a-b, a coma aberration was imposed through an SLM on the pump laser. For c-d, instead, second-order astigmatism was applied on the pump laser. Panels e and g show the reconstructed phase and amplitude distributions in $z=0$. In contrast, f and h show the intensity obtained by numerically propagating the fields in e and g to $z=19$ cm (to be compared with panels b and d, respectively). The phase distributions in e and g are shown with the tip and tilt contributions removed since these are due only to the imperfect centring of the intensity patterns in the two planes.}
    \label{fig:aberr}
\end{figure*}
We observed the free space propagation of the pump and phasematching contribution of an SPDC state generated in a 0.5 mm thick Type-I BBO crystal pumped by a pulsed 405 nm laser beam (pulse duration $150$ fs, repetition rate $80$ MHz). A Liquid Crystal Spatial Light Modulator (LC-SLM) was used to shape the spatial structure of the pump laser. A conceptual scheme of the experiment is sketched in Figure \ref{fig:setup} a, and a detailed setup is described in the Methods. Idler and signal photons can be spatially separated -- with 50\% probability-- by a non-polarising beamsplitter and imaged on two regions of a time-stamping camera (Timepix3D) \cite{nomerotski2019imaging, vidyapin2023characterisation, nomerotski2023intensified}. If the spatial distributions of the idler and signal do not overlap on the camera sensor, one can choose two regions of interest (ROI), defined as the groups of pixels hit by the idler and signal photons, respectively. The time-stamping resolution of the Timepix3D is less than $10$~ns and allows to extract coincidences between pixels contained in the ROIs, $\mathcal{C}(\mathbf{X}_i,\mathbf{X_s})$. From the 4-dimensional coincidence distribution is possible to extract marginals, e.g.~the spatial correlations along $x$:  $\mathcal{C}_x(x_i,x_s)=\sum_{y_i,y_s}\mathcal{C}(x_i,y_i,x_s,y_s)$. 
Alternatively, 2D sections of the 4D coincidence pattern can be extracted. As shown by Eqs.~\eqref{pump} and \eqref{phasematch}, this allows to obtain the intensity distributions $\abs{\mathcal{E}_p(\mathbf{X+c})}^2$ and $\abs{\mathcal{\phi}(\mathbf{X-c})}^2$, where the constant $\mathbf{c}$ can be chosen at will. To fully exploit the accumulated data, one can sum up the intensities obtained for each value of $\mathbf{c}$ after appropriately shifting each distribution by $\pm\mathbf{c}$. In this way, it is possible to achieve smooth reconstructions from data obtained with a few minutes of exposure (see Supplementary Materials, Figure S1). A first example of the reconstruction is shown in Fig.~\ref{fig:setup} b and c, for the case of a Gaussian pump. The analysis was carried out in two different propagation planes at distances $z_1=7.5$ cm and $z_2=34.5$ cm from the image plane of the crystal. The spatial correlations show the evolution from a spatially correlated state (signal and idler photon are localised in roughly the same transverse position) to a spatially anti-correlated state (signal and idler photons are localised in opposite transverse positions with respect to the biphoton state propagation axis). The correlations do not appear particularly sharp in these two planes. However, they show width of the order of a single pixel in both the crystal image plane and the corresponding Fourier plane. The extracted pump intensities display the expected Gaussian distribution (slightly narrower in $z_2$ due to a wavefront curvature of the pump laser on the crystal plane). The extracted phasematching intensity shows the formation of the characteristic SPDC cone in a collinear phasematching configuration. 

It is well known that measuring the intensity distribution of a coherent beam in two different propagation planes can be used to retrieve the phase. This non-interferometric phase retrieval approach was first proposed by Gerchberg and Saxton~\cite{gerhberg1972practical} for the case where the two intensity distributions are measured in conjugate planes (i.e. the corresponding fields are one the Fourier transform of the other). Unfortunately, doing this for an SPDC state implies that, for instance, the reconstructed phasematching intensity will have a width of one or a few pixels in the first recording plane (and a similar effect will be observed for the pump distribution in the far field). This is due to the fact that the spatial extension of pump and phasematching functions are extremely different on these two planes. One can then think of looking for a compromise and choose two intermediate planes sufficiently far away from each other but where the spatial extension of the pump and phasematching are of the same order. However, in these cases, the Gerhberg-Saxton (GS) --as well as variations like the Fienup algorithm~\cite{fienup1978reconstruction, fienup1982phase}-- tends to have a worse convergence and be sensitive to noise. This is acceptable for applications where a phase pattern is generated to achieve target intensities but is less practical for phase retrieval since the GS can easily converge to local minima characterized by strongly irregular phase patterns.

Due to these difficulties, we focused on retrieving the phase structure of phasematching and pump using optimization algorithms. However, these algorithms relied on some assumptions about the modal structure of the measured quantities. This is not a big issue since the SPDC physics for thin crystals has been extensively studied, and SPDC spatial mode structures are expected to follow the field continuity.

As a first step, we consider the phase retrieval of the phasematching function $\phi$. In Fig.~\ref{fig:phasematching}-a and b, we report the measured intensity $\abs{\phi(\mathbf{X},z)}^2$ in two different planes. $\phi(\mathbf{X},z)$ propagates essentially as a strongly diverging beam. One can expect --as also predicted by the thin-crystal theory of SPDC-- that the phase of $\phi$ in a given plane is quadratic: $\arg{\phi(\mathbf{X},z)}=\pi \mathbf{X}^2/(\lambda z)$. When applying this phase structure on the amplitude measured in $z_2$ and numerically propagating back to $z_1$, one obtains the intensity displayed in Fig.~\ref{fig:phasematching}-c, which is in good agreement with the experimental result. The reconstructed phasematching function is shown in Fig.~\ref{fig:phasematching}-c. The same phase structure was obtained via an optimization algorithm based on the decomposition of the phasematching function in orthogonal modes. We describe this process of reconstruction of the pump phase in detail below.  

Pump lasers prepared in spatial structures different from a single Gaussian have been considered, e.g.~to shape correlations in the Orbital Angular Momentum (OAM) degree of freedom~\cite{d2021full} or to control multimode Hong-Ou-Mandel bunching or anti-bunching~\cite{walborn2003multimode}. In the latter case, one considers a pump field that is an asymmetric function of one transverse coordinate. This is obtained by applying a $\pi$-phase jump on an input Gaussian beam. Figure~\ref{fig:hg10}-a shows the pump intensity contribution to the SPDC state in two different propagation planes. To retrieve the phase structure of the corresponding field, we consider its approximation to a finite superposition of paraxial modes $f_{\kappa}(\mathbf{X},z)$,
\begin{equation}
\mathcal{E}_p(\mathbf{X},z)\approx\sum_{\kappa}c_{\kappa}f_{\kappa}(\mathbf{X},z).
\label{eq:mode_decomposition}
\end{equation}
Here, $\kappa$ is, in general, a set of discrete indices. If the pump intensity in two different planes is $I_i(\mathbf{X})=\abs{\mathcal{E}_p(\mathbf{X},z_i)}^2$, with $i=1,2$, then the optimal modal decomposition can be found minimizing the functional 
\begin{align}
    \mathcal{L}[\{c_{\kappa}\}]:=&\iint\sum_i\Biggr|I_i(\mathbf{X})-\bigr|\sum_{\kappa}c_{\kappa}f_{\kappa}(\mathbf{X},z_i)\bigr|^2\Biggr|\,d^2X.
\label{eq:maxlike}
\end{align}
The choice of the modes set and the range of values for $\kappa$ varies case-by-case. Typical choices can be the Hermite-Gauss or Laguerre-Gauss sets \cite{siegman1986lasers}. However, spatially shaped beams are often generated by applying a phase pattern with line or point singularities on an input Gaussian beam. This typically requires a large number of coefficients in the decomposition equation~\eqref{eq:mode_decomposition}. In this situation, another convenient choice is the over-complete set of Hypergeometric-Gauss modes $\text{HyGG}_{-\abs{\ell},\ell}(\mathbf{X})$~\cite{karimi_07} -- the detailed expression used in this work is given in the Methods. We recall that the index $\ell$ refers to the Orbital Angular Momentum (in units of $\hbar$) carried by the corresponding mode. The coefficients $c_{\ell}$ will thus give the OAM spectrum of the analyzed field. Figure \ref{fig:hg10}-b shows the field intensity of the decomposition \eqref{eq:mode_decomposition} obtained minimising Eq.~\eqref{eq:maxlike} where $f_\kappa\rightarrow \text{HyGG}_{-\abs{\ell},\ell}$. The excellent agreement with the experimental data is also highlighted in Fig.~\ref{fig:hg10}-c, where the experimental amplitude distribution along $y$,  for fixed $x=1.9$ mm, is compared with the reconstructed one. Figure~\ref{fig:hg10}-c shows the reconstructed field in ${z=0}$. A phase variation of $\sim \pi$ between the two intensity maxima can be observed, together with an overall smooth quadratic phase due to the imperfect collimation of the pump on the crystal. In Fig.~\ref{fig:hg10}-e, the OAM power spectrum is reported, showing how the main contribution comes from two OAM modes with $\ell=\pm 1$. The smaller $\ell=0$ contribution (associated with a Gaussian mode) is due to a residual misalignment of the SLM phase mask. In Fig.~\ref{fig:oam4_2}, we consider a pump created in an unbalanced superposition of OAM modes. Again, a finite decomposition in  15 HyGG modes (with $\ell=-7, \ldots,7$) yields a reconstruction that matches well the experimental data.

The use of a finite set of orthogonal modes can be less efficient for fields with no phase singularities. In most applications SPDC is generated by a pump in the fundamental mode of the laser cavity, however, the actual phase and amplitude can be altered by imperfections in the experimental setup. Thus, one can expect that the pump contribution has a smooth phase factor that can be expanded in Zernike polynomials. We provide proof of principle for these applications by introducing specific optical aberrations on the Gaussian pump with a UV-SLM. Figure~\ref{fig:aberr} a-d shows the extracted pump shapes in the crystal image plane $z_1=0$ and at $z_2=19$ cm for cases in which a coma and a second order astigmatism phase was introduced on the pump.  We assumed the phase $\xi_p(\mathbf{X},z=0)$ to be a superposition $\xi_p(\mathbf{X},z=0)=\sum_{n,m}\gamma_{n,m}Z^{m}_{n}(\mathbf{X})$, where $Z^{m}_{n}(\mathbf{X})$ are Zernike polynomials \cite{born2013principles} and $\gamma_{n,m}$ are real coefficients, while the amplitude is given by the square root of the experimentally retrieved intensity. Due to the lack of analytical expression for the propagation of aberrated modes, we relied on the use of a genetic algorithm instead of the maximum likelihood approach used in the previous examples. Random choices of $\gamma_{n,m}$ were used to define the individuals that initialize the genetic algorithm. A numerical Fresnel propagation approach was used to calculate the fields in $z_2$ resulting from the different individuals and compare the intensity with the experimental one. The function in Eq.~\eqref{eq:maxlike} was here used as the Fitness function of the genetic algorithm (details are given in the Methods section and the Supplementary materials). Panels f and h of Fig.~\ref{fig:aberr} show the best-reconstructed pump field at $z_2$ --which is in good agreement with the experiment--, and the corresponding amplitude and phase in $z_1$ are shown in panels e and g.

\section{Discussion and Conclusions}
In conclusion, we have shown a powerful application of coincidence imaging of biphoton states. Spatially resolved second-order correlations allow the extraction of information about the two main physical contributions to the SPDC biphoton states, the spatial structure of the pump beam and the phasematching function, which is determined by the physical properties of the nonlinear crystal used for the biphoton state generation. The intensity of these two functions can be extracted at any distance from the crystal, and the relationship of the obtained intensities at different distances is given by an appropriate paraxial propagation. We exploited these results to extract, employing optimization methods, the phase of the two investigated functions and, thus, the full biphoton state. It must be stressed that this high-dimensional state reconstruction requires only two spatially resolved coincidence measurements, which, thanks to modern time-stamping cameras, can be performed in a few minutes without any control of the biphoton source. However, the renunciation of interferometric methods, and thus of direct phase measurements, comes at the expense of designing the proper algorithm to find the best phase structure that describes the experimental results. For smooth phases, if the field intensities are retrieved in propagation planes too close to each other, there will be a higher uncertainty in the reconstructed phase. Moreover, the separation of pump and phasematching can be rigorously achieved in the thin crystal limit, when the transverse walk-off between pump and down-converted photons is negligible. The more general scenario will require more robust approaches to analyze the 4D second-order correlations in different propagation planes and extract the full phase patterns. We expect that this problem could be tackled by utilizing properly trained Neural Networks. Lastly, the ability to separate pump and phasematching contributions opens new opportunities for quantum imaging applications, some of which will be explored in future works. 

\section{Methods}
\subsection{Experimental Setup}

\begin{figure}
\centering
\includegraphics[width=0.8\columnwidth]{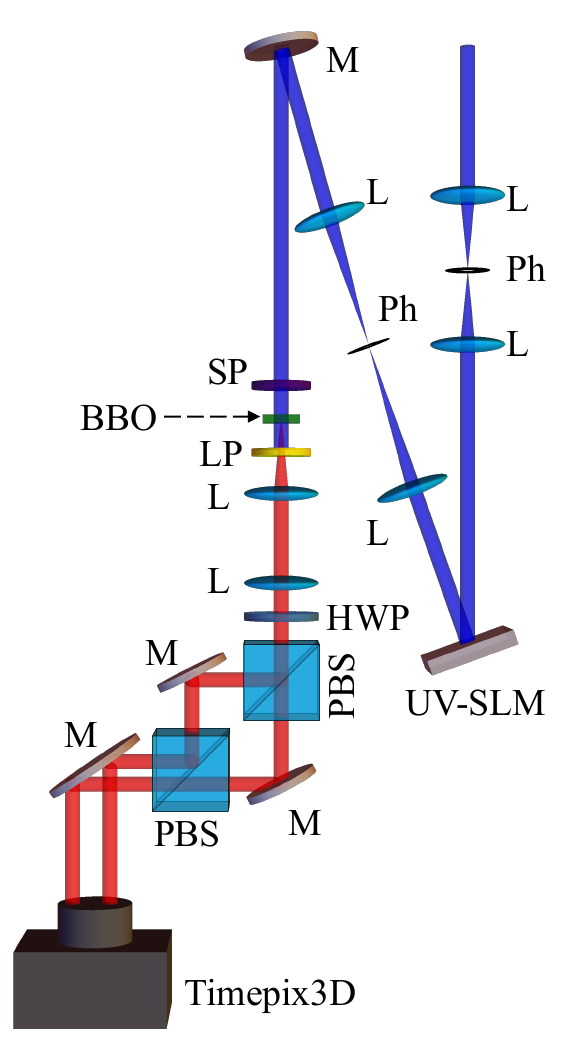}
\caption{{\bf Detailed Experimental Setup}, L: Lens, Ph: Pinhole, UV-SLM: Ultraviolet-Spatial Light Modulator, M: Mirror, SP: Short-Pass Filter, LP: Long-Pass Filter, HWP: Half Wave-Plate, PBS: Polarizing Beam Splitter}
    \label{fig:detailedsetup}
\end{figure}

The experimental setup is described in detail in Fig. \ref{fig:detailedsetup}. A 405 nm pump laser is generated as the second harmonic of a Ti:Sa pulsed laser (Chameleon Vision II). Then, a magnifying system of lenses with a pinhole in the beam's focus is used to generate the desired Gaussian beam that is then sent to the Spatial Light Modulator (SLM). The beam's phase and amplitude are structured using the amplitude-phase masking technique~\cite{bolduc2013exact}, which requires appropriate phase masks displayed on the SLM and selecting the first diffraction order. The latter is achieved by placing an iris in a de-magnifying system of lenses after the SLM. Sending the resulting structured pump beam through the BBO-Type I crystal, the SPDC is generated and collimated by a 75 mm lens. After the crystal, the pump beam is filtered by a low-pass filter. Signal and idler photons are separated with 50 \% probability, by a half wave-plate and a polarizing beamsplitter (PBS), instead of using a non-polarising beamsplitter. This approach guarantees that the two photons are now orthogonally polarised. Using two mirrors and a PBS, one can simultaneously match the optical paths of the two photons and propagate them along parallel paths, with a relative displacement smaller than the camera sensor. A bandpass filter (centred at 810 nm and with 10 nm bandwidth) is mounted on the camera intensifier to select frequency degenerate photons. The Timepix camera collects spatially resolved time stamps with $\sim$ 1~ns resolution~\cite{nomerotski2023intensified} from which the $4$D coincidence distribution is extracted.

\subsection{Hypergeometric-Gaussian modes}

Hypergeometric-Gaussian modes $\text{HyGG}_{p,\ell}$ with $p=-\abs{\ell}$ are a set of paraxial modes whose expression in the waist plane is given by $$\text{HyGG}_{-\abs{\ell},\ell}(r,\phi,z=0)\propto\exp(-r^2/w^2)\exp(i\ell\phi),$$ where $(r,\phi,z)$ are cylindrical coordinates, $\ell$ an integer number, and $w$ a real parameter corresponding to the waist radius. Here, we consider the more general case where a quadratic phase is added on the waist plane:
\begin{equation}
    \text{HyGG}_{-\abs{\ell},\ell}(r,\phi,z=0)\propto e^{-r^2/w^2}e^{-i\pi r^2/\lambda \mathcal{R}}e^{i\ell\phi},
\end{equation}
where $\mathcal{R}$ is the wavefront curvature radius. The expression for $\text{HyGG}_{-\abs{\ell},\ell}(r,\phi,z)$ for arbitrary $z$ can be obtained analytically by calculating the Fresnel propagator. It is convenient to use dimensionless coordinates: ${\rho=r/w}$, ${z=\zeta/z_0}$ (with ${z_0=\pi w^2/\lambda}$) and ${\xi=\mathcal{R}\lambda/\pi w}$. The Fresnel integral thus reads
\begin{align}
    \text{HyGG}_{-\abs{\ell},\ell}&(r,\phi,z)\propto\iint e^{-\rho'^2(1-i/\xi)}e^{i\ell\phi'}\times \cr &e^{-i[\rho'^2+\rho^2-2\rho\rho'\cos(\phi-\phi')]/\zeta}\rho'd\rho'd\phi'\cr=&e^{-i(\frac{\rho^2}{\zeta}-\ell\phi)}\times\cr &\int_0^{\infty}\rho'd\rho'J_{\abs{\ell}}(\frac{2\rho\rho'}{\zeta})
e^{-(\frac{\sigma+i}{\sigma})\rho'^2}
\end{align}
where $J_{\abs{\ell}}(.)$ is a cylindrical Bessel Function of order $\abs{\ell}$ and $1/\sigma:=1/\xi-1/\zeta$. The solution to the last integral (see Ref.~\cite{luke2014integrals}) can be expressed in terms of modified Bessel functions:
\begin{align}
    \int_0^{\infty}&\rho'd\rho'J_{\abs{\ell}}(\alpha\rho') e^{-\beta\rho'^2}=\cr&\frac{\sqrt{\pi}\alpha e^{-\frac{\alpha^2}{8\beta}}}{8\beta^{3/2}}[I_{\frac{\abs{\ell}-1}{2}}(\frac{\alpha^2}{8\beta})-I_{\frac{\abs{\ell}+1}{2}}(\frac{\alpha^2}{8\beta})],
\end{align}
with $\alpha:=2\rho/\zeta$ and $\beta:=\frac{\sigma+1}{\sigma}=1+i(1/\zeta-1/\xi)$. The functions $I_{\frac{\abs{\ell}\pm1}{2}}$ are modified Bessel functions of the first kind.

\subsection{Details of genetic algorithm}
Genetic algorithms evolve a population of candidate solutions toward optimal solutions to the problem of minimizing a cost function. In our case, the cost function is given by Eq.~\eqref{eq:maxlike}. The analytical expressions for the propagation of families of optical modes at a finite distance $z$, such as Hermite-Gauss or Laguerre-Gauss modes, enable the adoption of standard minimization routines. This is not possible in the case of aberrated beams, where there is no analytical formula for the propagated field. Genetic algorithms offer a suitable alternative, as these iterate from random guesses that evolve to the physical solutions of the optimization problem~\cite{holland:92}. In the following, we detail the sequence of operators used in our genetic algorithm. An initial population of $N$ individuals is randomly generated within the range $[-20,20]$, where each individual is a set of real coefficients $\gamma$ (\emph{genes}) for the chosen Zernike polynomials. In groups of $k$, these individuals compete for the possibility to reproduce. Only the individual better minimizing the cost function within each pool is given access to the reproduction stage. This is the so-called \emph{tournament} mechanism, working as a selection operator~\cite{goldberg:91}. The reproduction occurs in the form of \emph{blend crossover}, largely employed to mate real-valued individuals~\cite{eshelman:93}. When two individuals reproduce, two newborns originate as a weighted mixture of the parents. To emulate the mutation of individuals in a natural environment, genetic mutations are included in the workflow as Gaussian noise with mean $\mu$ and standard deviation $\sigma$, potentially affecting each gene of newborn individuals~\cite{kramer:17}. Blend crossover and mutation are non-deterministic operators, occurring with probability ${p_c}$ and ${p_m}$, respectively. To push the algorithm to a faster convergence, our algorithm is equipped with \emph{elitism}, i.e., the best individual from the parent generation is guaranteed a place in the next one, replacing the worst individual of the offspring. The algorithm ends when a certain condition is verified~\cite{yao:93}. In our implementation, the maximum number of generations $N_\text{gen}$ is adopted as termination criterion. The user defines by hand all the parameters determining the evolutionary sequence (also called \emph{hyperparameters}). The complete set of hyperparameters used in our algorithm is listed in Table~\ref{tab:GA_parameters}. 
\begin{table}[!h]
\caption{Genetic algorithm hyperparameters}
\centering
\renewcommand{\arraystretch}{1.5} 
\begin{tabular}{ll}
\hline
\hline
Population size       & $N = 100$                     \\
Number of generations & $N_\text{gen} = 30$               \\ 
Tournament size       & $k = 4$     \\
Crossover probability & $p_c = 0.9$                \\
Mutation probability  & $p_m = 0.04$                \\
Gaussian mutation     & $\mu = 0$, $\sigma = 0.5$ \\
\hline
\hline
\end{tabular}
\label{tab:GA_parameters}
\end{table}

\section{acknowledgment}
This work was supported by the Canada Research Chair (CRC) Program, NRC-uOttawa Joint Centre for Extreme Quantum Photonics (JCEP) via the Quantum Sensors Challenge Program at the National Research Council of Canada, and Quantum Enhanced Sensing and Imaging (QuEnSI) Alliance Consortia Quantum grant.

\bibliography{bibliography.bib}

\vspace{1 EM}

\noindent\textbf{Author Contributions}
A.D. conceived the idea.  N.D. and A.D. performed the experiment. N.D., F. D. C. and A.D. prepared and tested the phase retrieval algorithms.   E.K. supervised the project. N.D. and A.D. prepared the first version of the manuscript. All authors contributed to the writing of the manuscript.
\vspace{1 EM}

\noindent\textbf{Data availability}
\noindent
The data that support the findings of this study are available from the corresponding author upon reasonable request.
\vspace{1 EM}

\noindent\textbf{Code avialability}
\noindent
The code used for the data analysis is available  from the corresponding author upon reasonable request.

\vspace{1 EM}
\noindent\textbf{Ethics declarations} Competing Interests. The authors declare no competing interests.

\vspace{1 EM}
\noindent\textbf{Corresponding authors}
Correspondence and requests for materials should be addressed to aderrico@uottawa.ca.

\end{document}


\clearpage
\onecolumngrid
\renewcommand{\figurename}{\textbf{Figure}}
\setcounter{figure}{0} \renewcommand{\thefigure}{\textbf{S{\arabic{figure}}}}
\setcounter{table}{0} \renewcommand{\thetable}{S\arabic{table}}
\setcounter{section}{0} \renewcommand{\thesection}{S\arabic{section}}
\setcounter{equation}{0} \renewcommand{\theequation}{S\arabic{equation}}
\onecolumngrid

\begin{center}
{\Large Supplementary Material for: \\ Biphoton State Reconstruction via Phase Retrieval Methods}
\end{center}
\vspace{1 EM}

\subsection*{Data analysis}

The extraction of phasematching and pump intensity at a given plane is illustrated in more detail in Fig.~\ref{fig:data_analysis}. After the coincidence distributions are recorded, one can postselect the coincidences corresponding to the green bands in the correlation space shown in Fig. \ref{fig:data_analysis} (first two rows). Depending on whether one wants to extract the pump or the phasematching, the selection regions are inclined at $-45°$ or $+45°$ in the correlation space, respectively. The coincidence images resulting from this selection correspond to the desired intensities. As shown in the third row, these intensity patterns are centered in different positions depending on the specific choice of the selective band. Moreover, the narrow selection implies an overall low number of counts so that the coincidence images for a single are noisy. However, by correcting for the shift and accumulating the coincidence images corresponding to all the possible choices of postselection one can achieve a high-quality result from a dataset obtained from a few minutes of exposure (see examples in the last rows).

\subsection*{Convergence of the genetic algorithm to the best solution}

In Fig.~\ref{fig:Supp_GA}  additional details of the genetic algorithm results are shown. We report the best individual intensities for some intermediate generations showing the rapid convergence to the experimental data. The comparison with the experiment is well quantified by the similarity $S$ defined as \begin{equation}
S=\frac{\sum_{x,y}\sqrt{I_{exp}(x,y) I_{rec}(x,y)}}{\sqrt{\sum_{x',y'}I_{exp}(x',y')\sum_{x'',y''}I_{rec}(x'',y'')}}
\end{equation}
where $I_{exp}$ is the experimentally retrieved intensity and $I_{rec}$ is the best output of each generation in the genetic algorithm. Note that $S=1$ for identical distributions. In our case, $S$ converges to values close to 0.9. The residual mismatch is in good part due to noise in the experimental data (which should not be reproduced by the numerical propagation).

\subsection*{Supplementary Figures}

\begin{figure}
\includegraphics[width=\textwidth]{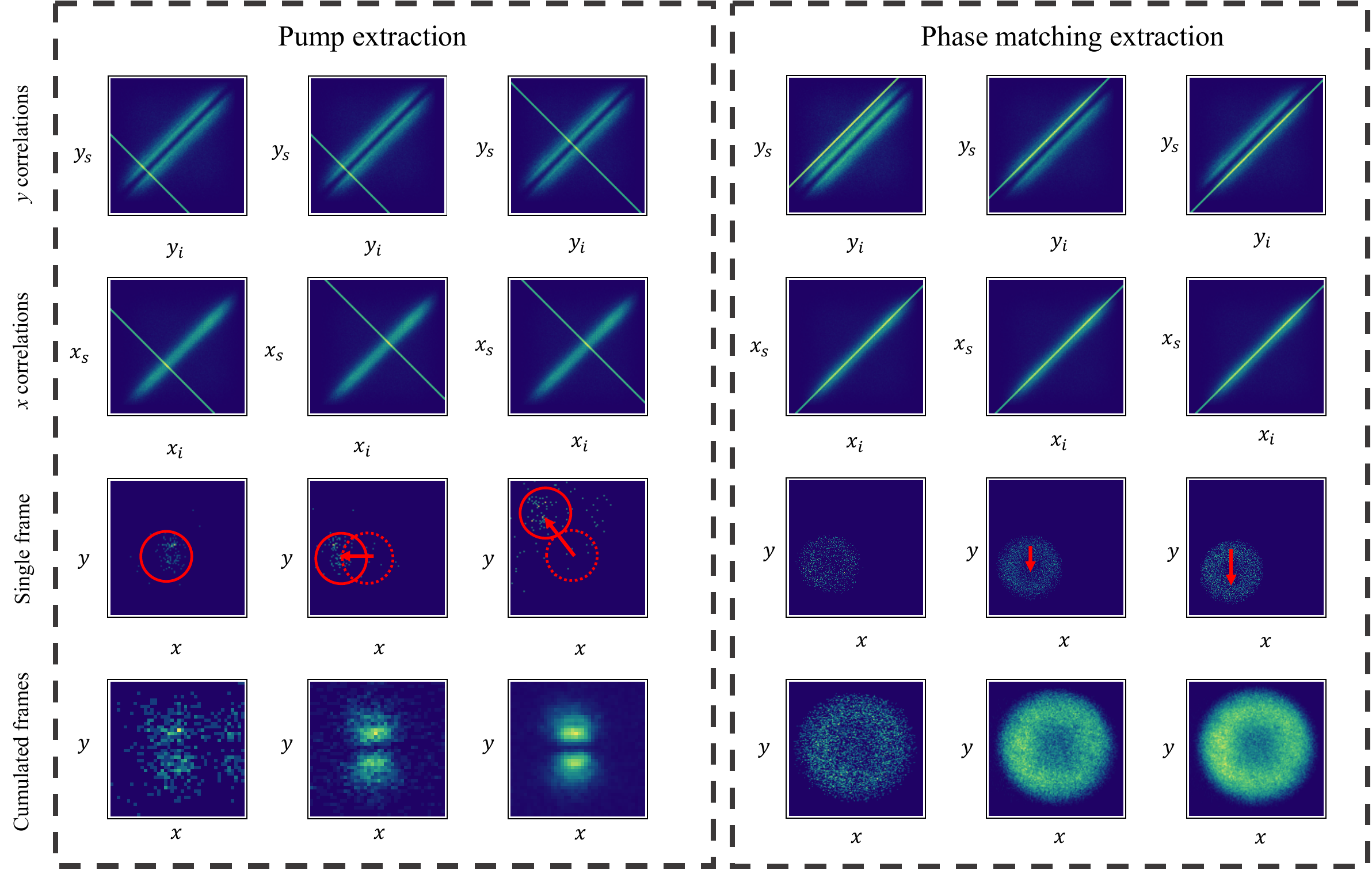}
    \caption{{\bf Illustration of pump and phasematching function extraction}. The process of pump and phasematching intensity extraction is illustrated. The first two rows show the spatial correlations with a 1-pixel wide green band which indicates the region of coincidence postselection. The third row shows the coincidence distribution after the considered postselection. It is possible to observe how the spatial shift of this distribution is affected by the position of the postselection region. After compensating for this shift the results can be accumulated to obtain a less noisy distribution of the pump and phasematching intensity, as shown in the last row.     }
    \label{fig:data_analysis}
\end{figure}

\begin{figure}
\includegraphics[width=\textwidth]{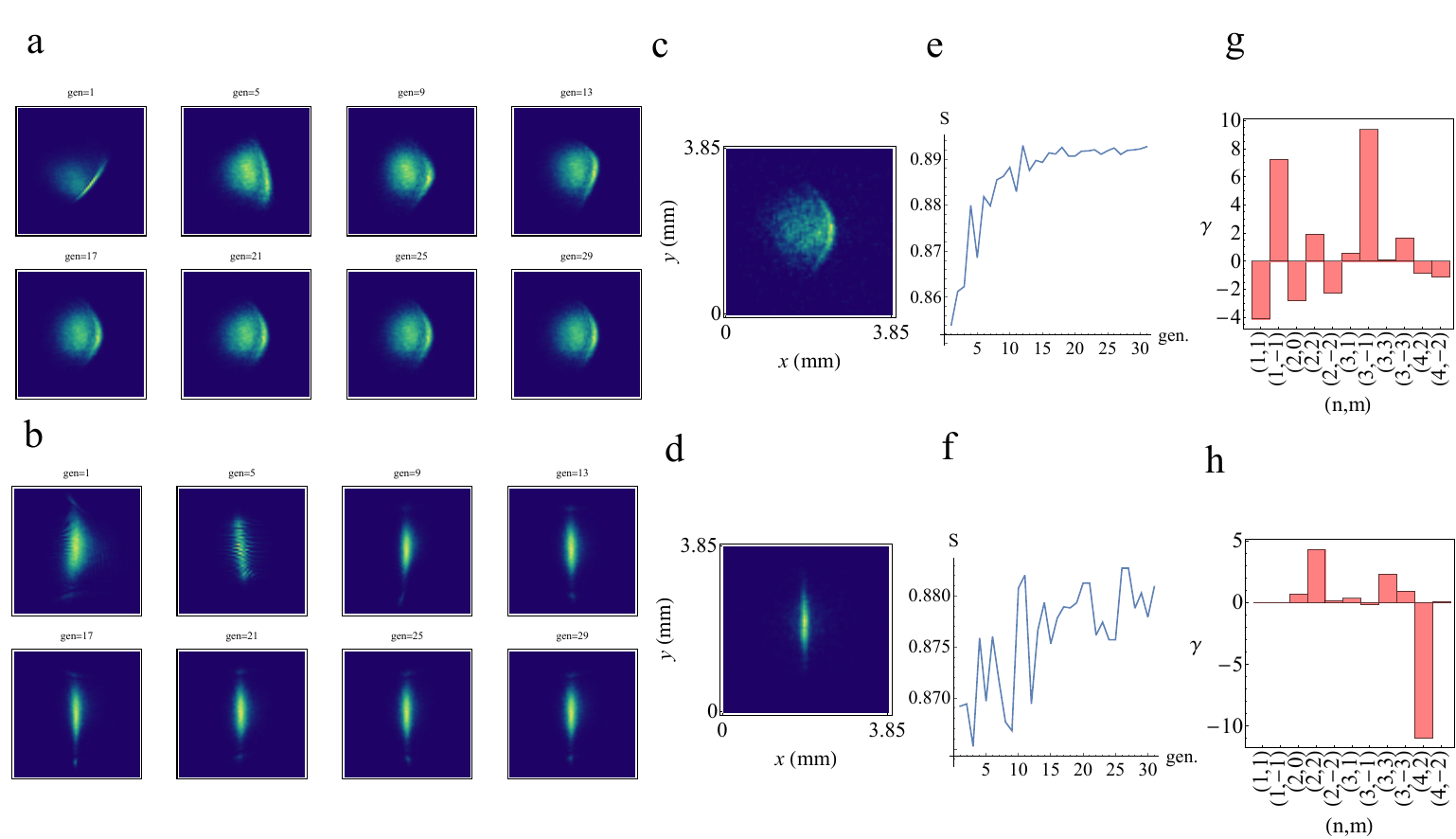}
    \caption{{\bf Detailed results of Genetic algorithm}.  Intensities of the best individual at each generation for a, coma, and b, second-order astigmatism. c and d show, respectively, the experimental results used as a target of the genetic algorithm. In panels e and f we show the evolution of the similarity $S$ between the best individual and the target intensity. Panels g and h show the Zernike decomposition corresponding to the best individual of the last generation.   }
    \label{fig:Supp_GA}
\end{figure}

Figure~\ref{fig:data_analysis} illustrates the data analysis process. Figure~\ref{fig:Supp_GA} shows the performance of the genetic algorithm and the retrieved Zernike decomposition.
\newpage

\bibliography{bibliography.bib}